# Tourism destinations as digital business ecosystems

Rodolfo Baggio[a] and Giacomo Del Chiappa[b]

[a] Master in Economics and Tourism and
Dondena Center for Research on Social Dynamics
Bocconi University, Italy
rodolfo.baggio@unibocconi.it

[b] Department of Economics and Business
University of Sassari and CRENoS, Italy
gdelchiappa@uniss.it

**Abstract**

Tourism has been experiencing very relevant changes since when Information and Communication Technologies (ICTs), in all their forms, have started to pervade the industry and the market. In the last decade, a new concept gained the attention of both researchers and practitioners, that of Digital Business Ecosystem (DBE). It can be considered as a technological infrastructure aimed at creating a digital environment to support and enhance networking between enterprises and stakeholders operating within a sector. Aim of this paper is to assess the extent to which the technological connection has affected the structural configuration of the tourism system and, specifically, of tourism destinations. The present study argues that two components can be considered when assessing the relationships among stakeholders within a tourism destination: a real and a virtual one. Further it shows how these two components are structurally strongly coupled and co-evolve forming a single system.

## 1 Introduction

In the last decades, Information and Communication Technologies (ICTs) have radically and unforeseeably changed society as a whole. New ways of collective human behaviour have appeared and individuals, society, and ICTs are today so deeply intertwined in a dynamic feedback process that a profound restructuring in the whole of human activities has occurred. Rather obviously, travel and tourism, as activities deeply rooted in human nature, have been retransformed as well, and the nature of the entire sector has been (and is still being) deeply modified. ICTs and travel and tourism have developed, since the beginning of their recent history, a strong relationship. The first ever industrial real-time computerized system is an airline reservation system (Sabre) and appeared in the early 1960s. Since then, Internet, ICTs and the so-called Web 2.0, have transformed the structure of the market value chain, altered the power position of stakeholders and generated opportunities and threats for all organisations involved in the tourism system (Berne et al., 2012; Buhalis & Law, 2008: Del Chiappa, 2013).

As well noted by the seminal work of Werthner and Klein (1999: 1): "*Information technology does not only enable, but also induces changes*", mainly for activities that rely so extensively on information exchanges such as travel and tourism. Broadly, it could be argued that with the World Wide Web commercial and business functions have been developed to a good level of sophistication thus making real the idea of a networked organization able to function without spatial or temporal constraints. Furthermore, digital marketing channels are impacting operational practices of firms, their functional structure and the way they operate in a globalised economic environment (CMO Council, 2011). This is particularly relevant in fragmented sector, such as tourism, where ICTs may allow small and medium enterprises (SMEs) to be flexible and efficient without suffering from market fluctuations, despite the disadvantages due to their size (Dini et al., 2008).

One question that arises today is: is there anything beyond what might be called an *anecdotal* evidence for the importance of the role played by ICTs in tourism? Is there some indication that this strong relationship is, or has become, deeper?

Aim of this paper is to examine this question by adopting an uncommon perspective, and assess the extent to which the technological connection has affected the structural configuration of the tourism system. We shall consider a tourism destination, the essential unit of study for understanding the phenomenon, and study the network formed by its physical and virtual components. When relationships are strong, it is natural to call for a concept such as the one of Digital Business Ecosystem (DBE), that can offer a different view for understanding the structural and dynamic behaviour of our object of study.

The paper is structured as follows. Section 2 introduces the concept of DBE and briefly discusses its application in the tourism field along with a short discussion on the main literature dealing with the analysis of coupled networks. Section 3 presents the methods used in this paper and section 4 discusses the outcomes of the analysis and the main implications. Finally, section 5 concludes the paper summarizing what presented and highlighting possible future developments for this line of investigation.

## 2 Digital business ecosystems

We can restate the development of modern ICTs observing its evolution from a simple tool to improve the efficiency of some task by automating operations to a complex system which plays a crucial role affecting the very essence of business processes not only from an operational point of view, but also, and more importantly, from a strategic point of view.

If we consider only the recent history (Nachira, 2002) we started from having available simple functions to exchange messages (e-mail). Then a new form of *mass communication* appeared. The World Wide Web has allowed unprecedented possibilities to make easily and cheaply available a wealth of materials to a wide and undifferentiated (in time and space) audience. As a consequence, commercial and business functions have been developed to a good level of sophistication so that the idea of a networked organization has become a reality, easing the capability to conduct business without having to be constrained by spatial or temporal factors.

The progress to a higher socialization of ICTs has now made much more relevant (and fashionable) the concept of digital business ecosystems. At the very beginning the concept was not well delineated and defined. The concept obtained a broad definition in the framework of a EU funded project (Nachira, 2002; Nachira et al., 2007). As reported (Nachira et al., 2007: 5) "The synthesis of the concept of Digital Business Ecosystem emerged in 2002 by adding *digital* in front of Moore's (1996) *business ecosystem* in the Unit ICT for Business of the Directorate General Information Society of the European Commission".

The analogy used is the one with a natural ecosystem, the biological community of interacting organisms fully embedded in their physical environment. Thus, a DBE is a networked system which comprises the buyers, suppliers and makers of certain products or services, the socio-economic environment, including the institutional and regulatory framework (the business ecosystem defined by Moore, 1996) complemented by a technological infrastructure aimed at creating a digital environment for the networked organizations that supports the cooperation, the knowledge sharing, the development of open and adaptive technologies and evolutionary business models (Stanley & Briscoe, 2010). In others words, a digital ecosystem is a transparent virtual environment where open relationships between entities are established thus determining interaction and knowledge sharing, and where each entity is committed and cooperative (Boley & Chang, 2007). In a digital ecosystem "the network can be physical and logistical or virtual, can be local or global, or a combination of all the above" (Nachira et al., 2007: 8). The leadership structure is dynamic and may be formed and dissolved in response to any stimulus coming from the environment. Further, DBEs oscillate between multiple stable states without having a single optimal or equilibrium configuration (Salmi, 2001).

By its very nature a DBE is a complex adaptive system that exhibits properties of self-organization, scalability, dynamic adaptation to the environment (Baggio, 2008). In a DBE it is possible to recognize two main components: a *physical* one, composed of the business stakeholders in a certain economic or industrial sector and its *virtual* complement formed by the technological equivalents of these stakeholders. The two components are structurally strongly coupled and co-evolve forming a single system. The real part generates the virtual one, but, given the strong relationship between the two, all modifications, changes or perturbations originating in one of them rapidly propagate to the whole DBE (see section 2.2). The interactions within the combined network can be harmonised via ICTs or other traditional forms of coordination mechanism (face-to-face or technology mediated), thus confirming the idea that the offline and online worlds should be taken into account together when analysing a DBE (Dini et al., 2008).

Digital Ecosystems have been considered highly relevant especially in the case of highly fragmented sectors where a high number of SMEs are operating, as it is in the case of tourism. Indeed, in this circumstances, DBEs are considered being able to promote content sharing and Business-to-Business (B2B) interactions thus helping formation of dynamic, efficient and self-organising networks (Dini et al., 2008), to produce opportunities to form alliances and thrive in the network (Moore, 1993) and, finally, to expand the innovation ecosystem outside the firm boundaries thus enhancing the overall competitiveness (Karakas, 2009).

**2.1 Tourism DBEs**

Strangely enough, despite the vast literature on the crucial role ICTs have for the contemporary tourism industry, very little research can be found on the topic of digital business ecosystems in the tourism field. The term seems to be more a fashionable way used by popular press to describe the strong relationship between tourism and ICTs rather than a lens through which to examine the structure and the behaviour of a tourism system.

The DBE perspective seems to be a promising and interesting topic to be investigated in the tourism sector as a whole, and in tourism destinations in particular. Based on existing research, a tourism destination may be considered as a cluster of interrelated stakeholders (both public and private) embedded in a social network (Baggio et al., 2010b). In such a network, an individual company's performance depends also on the behaviour of other companies and vice versa (Freeman, 1984; Del Chiappa & Presenza, 2012). Further, the performance of a tourism destination as a whole depends on the web of connections between the various players and not only on the intrinsic characteristics of the destination (March & Wilkinson, 2009). That said, it appears that the DBE and its support in enhancing network interactions can be pivotal for destination competitiveness.

The present work aims at exploring this somewhat neglected area of tourism research carrying out an empirical investigation in two tourism destinations by assuming that two components need to be considered at the same time: the real and the virtual one.

**2.2 A digression on coupled networks**

Network science has provided in the last years numerous tools for studying the structure and the dynamic behaviour of many complex systems present in nature, technology and society. Most studies have so far dealt with networks where vertices correspond to single elements or subsystems, and edges indicate interactions or relationships between vertices (da Fontoura Costa et al., 2011). However, a significant number of systems can be treated, more appropriately, as composite assemblies of interacting networks. Networks of different types, in fact, may combine in multiple ways and generate systems whose properties cannot be simply inferred by combining those of their constituents.

Saumell-Mendiola et al. (2012), for example, analyse epidemic spreading on interconnected networks and show that two networks well below their respective epidemic thresholds may sustain an endemic state when coupling connections are added, even in small number. Dickinson et al. (2012), find that in strongly coupled networks, epidemics occur across the entire system when a critical infection strength is overcome, while weakly-coupled systems exhibit mixed phases where an epidemic may occur in one network without spreading to the whole coupled system. Yağan et al. (2011) and Qian et al. (2012) study information spread in online social networks coupled to a physical network (made of firms, for example). They find that even if there is no full diffusion in the individual networks, an information epidemic can take place in the conjoint social-physical network.

Other authors examine the robustness of composite networks (Buldyrev et al., 2010; Vespignani, 2010). The failure of nodes in one network can lead to the failure of nodes in a coupled network that in turn can cause the escalation of failures in the first network, eventually leading to a complete disruption of the system. One consequence is that the value of the critical threshold is smaller than in an isolated network, suggesting that a collapse of the system will happen at a smaller level of sustained damage. More importantly, in interdependent networks the fragmentation occurs with an abrupt transition. This makes complete system breakdown even more difficult to anticipate or control than in a single network.

## 3   Materials and methods

Two Italian destinations are used here to assess the structural composition of the tourism DBE. One is the island of Elba, a known marine destination whose main networked characteristics have been deeply analysed elsewhere (Baggio, 2007; Baggio et al., 2010a; da Fontoura Costa & Baggio, 2009). The second is Livigno, a mountain area studied by Mulas (2010). For both destinations the networks of core tourism stakeholders were assembled together with those formed by their websites. In these networks the links between the different actors were uncovered following the methods extensively described in Baggio et al. (2010a). In both cases the networked elements were classified into two main categories: physical elements, representing the "real" companies and organizations, and virtual elements, the websites belonging to the tourism stakeholders.

A first analysis was conducted in order to assess the self-organization characteristics of the two networks. The method chosen consists of finding, with a stochastic algorithm, the communities that arise from the distribution of the linkages among all the elements in the networks. The communities (or modules) are groups of nodes more densely connected between them than with other nodes in the network. A modularity index measures the goodness of the division in groups; it is defined as:

$$Q = \sum_i (e_{ii} - a_i)^2 \qquad (1)$$

where $e_{ii}$ is the fraction of edges in the network between the nodes in group $i$, and $a_i$ the total fraction of links originating from the group and connecting nodes belonging to different ones. In other words, $Q$ is the fraction of all links that lie within a community minus the expected value of the same quantity that could be found in a graph having nodes with the same degrees but with a random distribution of the links. The index is always smaller than one; higher values indicate better separations of the communities. For easing the comparison between different networks with different numbers of communities, the index can be normalized by the number of modules $m$ (Du et al., 2009).

$$Q_{norm} = \frac{m}{m-1} Q \qquad (2)$$

In the last years a wealth of possible techniques have been put forward and employed for detecting communities (for a thorough review see Fortunato, 2010). Here we chose a recent proposal by Karrer and Newman (2011). They use a modified version of blockmodelling for detecting the community structure in a network. The goal of blockmodelling is to reduce a large network to a smaller structure that can be interpreted more easily. It is an empirical procedure centred on the idea that nodes in a network can be grouped according to the extent to which they exhibit some form of structural equivalence (Doreian et al., 2004). Usually the algorithm starts with some specified blockmodel. The solution is then found by iteratively changing the modules' compositions until a criterion function is minimized. As Karrer and Newman note, however (2011: 1): "*most blockmodels, however, ignore*

*variation in vertex degree, making them unsuitable for applications to real-world networks, which typically display broad degree distributions that can significantly affect the results*". They use, therefore, a modified algorithm which takes into account the real degree distribution of the network analysed and show how the results obtained are greatly significant in highlighting the structural characteristics of the system that arise, independently from the nature of the components.

Once identified the communities in our networks we measured, for each module, the proportion of nodes representing the physical and the virtual components in order to assess the interrelation possibly present between them.

The second investigation concerns the efficiency of the digital ecosystem compared with the one of the pure physical component. To this aim, a *cost* was assigned to each links. Specifically, three different values were used: 1 for a link between two virtual elements, 2 for a link between a virtual and a physical element and 3 for a link between two physical elements. Although arbitrarily chosen, these values can reasonably represent the real-life efforts in establishing and maintaining such connections, as the analyses on transaction costs for real and virtual connections and operations has shown (Hagel & Armstrong, 1997; Rayport & Sviokla, 1995; Upton & McAfee, 1996). The efficiency of the weighted network are calculated at global and local level (da Fontoura Costa et al., 2007). They measure the capability of the whole system (global efficiency: $E_{Glob}$) or of a single node (local efficiency: $E_{Loc}$) to allow for exchanges (information, goods etc.). Network efficiencies depend strongly on the general topology of the network (number and distribution of connections), and are obviously influenced by the cost associated with each connection.

## 4  Results and discussion

The two networks examined show topological characteristics that clearly indicate their complex and heterogeneous structure (Baggio et al., 2010a; Mulas, 2010). This fact, as known, has significant effects on the dynamic behaviour of the system and on the processes that unfold over these networks, such as information diffusion and spreading, robustness or fragility, or self-organization in modular components (da Fontoura Costa et al., 2011; Newman, 2010). In particular, the distribution of the connections each node in the network has (termed degree distribution) exhibits a marked scale-free structure (power-law form degree distribution), a well-known signature of complexity. Moreover, this topology is almost identical (apart from some scaling constant) for both the physical and the virtual components of the tourism systems (see Fig. 1 which shows the cumulative degree distributions for the networks studied) .

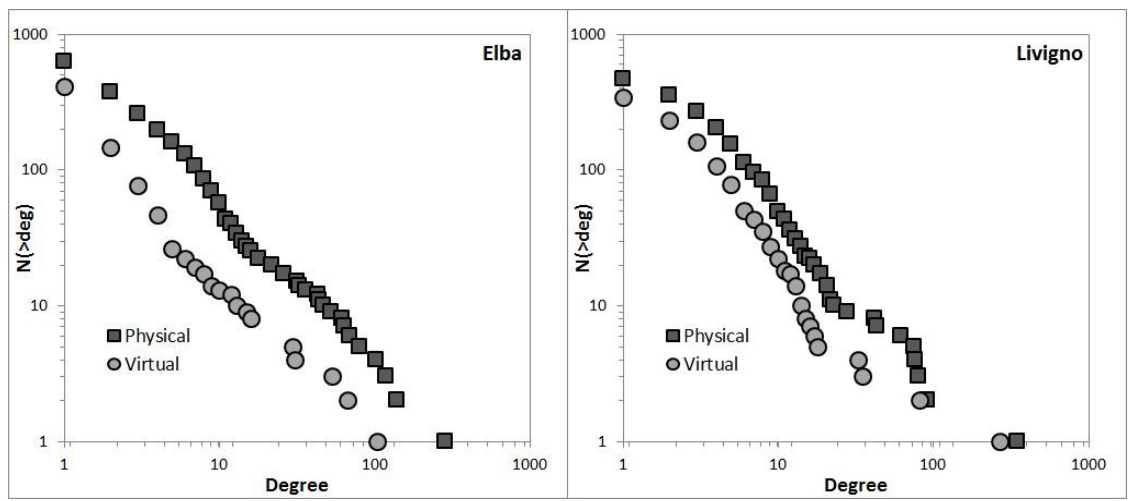

**Fig. 1.** Cumulative degree distributions for the physical and the virtual components of the Elba and Livigno networks

The modularity analysis recognises seven communities for the Elba network and eleven for Livigno. The normalised modularity index is $Q_{norm} = 0.1$ for Elba and $Q_{norm} = 0.5$ for Livigno, showing a much better separation of the latter's modules. This can be interpreted as due to a higher propensity to form cooperative groups by the Livigno's tourism operators. If we identify the nodes of these communities as belonging to the physical or the virtual components we obtain the situation depicted in Fig. 2. As can be seen, all modules have a mixed population and the distribution of both types of elements can be assumed to be rather uniform. On the average, a community in the Elba network has 48% of virtual elements and a Livigno community has 43%. The Gini coefficient, showing the uniformity of these proportions across all modules is 0.1 for Livigno and 0.2 for Elba (the coefficient is 0 for maximum uniformity, 1 for maximum inequality).

The first conclusion is therefore that from a structural point of view, the physical and the virtual components cannot be easily separated thus strongly reinforcing the idea that a DBE is more than just an anecdotic phenomenon. That said, it can be argued that the role of the virtual elements has become so important that they modify the very nature of the tourism systems considered.

Once ascertained the fundamental structural role of the virtual elements in a tourism destination, a study of the differences in the efficiency with which a network behaves when considered in its pure physical component or as an integrated real-virtual system can provide a stronger argument in favour of considering a DBE as such and not as a simple "addition" of two separate components.

In both our cases we calculated both the global and the local (individual) network efficiencies for the whole ecosystem and for the pure physical component. To make the analysis more realistic we considered, as stated in section 3, the "costs" of establishing and maintaining the relationship between different typologies. This analysis highlights well the contribution at all levels (for the whole system and for the single stakeholders) of the structural modifications that the introduction of technological elements provides.

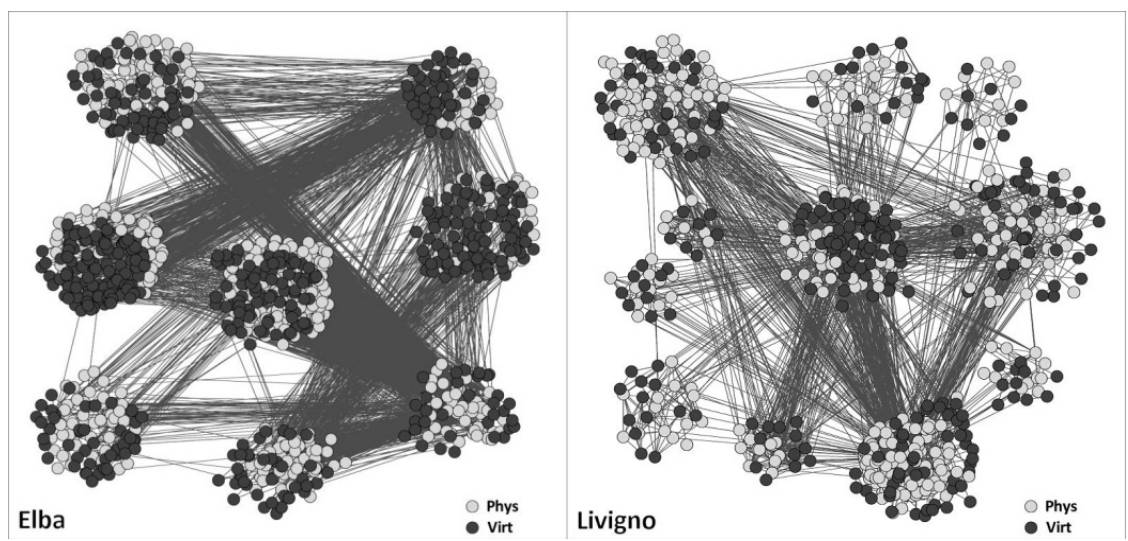

**Fig. 2.** The communities recognized by modularity analysis. Physical and virtual elements are identified

Table 1 reports the global efficiency coefficients for the cases examined. It is rather clear how the addition of the virtual component has a positive effect on the whole ecosystem.

**Table 1.** Global efficiency values

| Component | Elba | Livigno |
| --- | --- | --- |
| Physical | 0.118 | 0.144 |
| Ecosystem | 0.154 | 0.188 |
| *Difference* | *31%* | *30%* |

Fig. 3 shows the cumulative distributions of the local efficiencies for the two networks. The case of the whole ecosystem and the one of the pure physical component are highlighted. Given the highly non-normal shape of the distributions, comparing means would be scarcely meaningful. Visually, the difference between the two cases is clear for both systems. In order to assess the significance of this difference, a number of non-parametric tests can be run. In our case the Wilcoxon signed ranks, the 2-sample Kolmogorov-Smirnov and the marginal homogeneity tests seem relevant (Sheskin, 2000; Siegel & Castellan, 1988). Table 2 reports the results; all of them confirm the visual impression with very high significance.

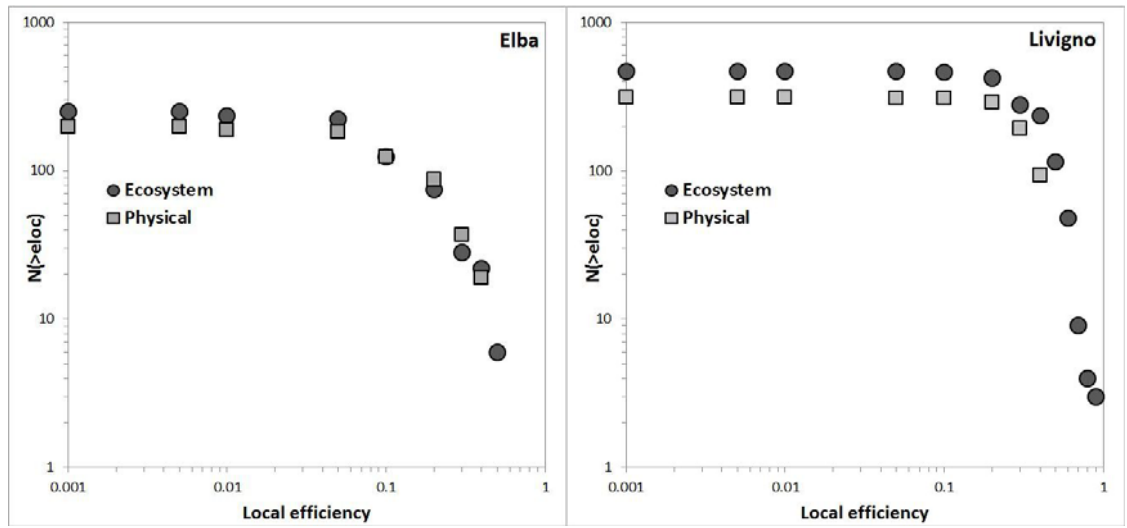

**Fig. 3.** Cumulative distributions for the local efficiencies

**Table 2.** Test results on the local efficiency distributions

| Test | Values | Elba | Livigno |
|---|---|---|---|
| Wilcoxon Signed Ranks | Z | -2.697 | -3.085 |
| | p-value (2-tailed) | 0.007 | 0.002 |
| 2-sample Kolmogorov-Smirnov | F | 0.105 | 0.209 |
| | p-value (2-tailed) | $<10^{-4}$ | $<10^{-11}$ |
| Marginal Homogeneity | Std. MH Statistic | -2.515 | 3.730 |
| | p-value (2-tailed) | 0.012 | $<10^{-4}$ |

All our initial hypotheses have thus been confirmed: the virtual component of a destination is a structurally crucial element, and its role is quite important in its effects on the dynamic behaviour of the system. Therefore the idea of considering a tourism destination as an integrated digital business ecosystem is not just a fashionable way of describing what happens today in the industry, but reflects a real intrinsic characteristic.

## 5  Concluding remarks

The strong relationship existing between ICTs and tourism leads almost naturally to considering a tourism system as an integrated ensemble in which both a real physical component (the companies and organisations active in the field) and a virtual one (the digital representations of the physical elements) act in a strongly coupled way. The resulting networked system can be seen as a digital business ecosystem in which the structure and the dynamic behaviour are of peculiar nature.

Despite the vast literature on the crucial role ICTs have for the contemporary tourism industry, still little research exists that analyses digital business ecosystems in the tourism field. With this work, we investigate this somewhat neglected area of tourism research carrying out an empirical analysis of two Italian tourism destinations.

Findings revealed that the interrelationships between the real and the virtual world are so tight that it will be difficult, if not impossible, to consider them separately any more. The coupling has reached a stage where the two elements influence each other so deeply that the idea of a DBE is not only a fashionable way to describe a tourism destination, but reflects a real characteristic of the system.

Needless to say, the implications for both researchers and practitioners are important, as they have, at this point, not only a number of examples to demonstrate the importance of ICTs in their areas, but also a strongly theoretically based validation of what up to now could have been considered a "motivated feeling".

Specifically, this paper adds to the growing research which applies network analysis to study tourism destinations from a systemic point of view and suggests that both the real and virtual components need to be addressed when assessing interorganisational relationships. In fact, the virtual dimension has become a structurally crucial element, especially if the tourist area as a whole is characterised by a significant diffusion of technological instruments.

The outcomes assessing the strict relationship between the tourism destination networks that can be drawn based on real and virtual perspectives are relevant also for marketing practices. Indeed, they suggest that destination managers cannot treat the virtual world as a separate entity any more, but they should consider online activities not differently from all the other more traditional ones. Moreover, by favouring their diffusion and integrated usage, they could achieve a much better functioning of the system by increasing its efficiency both at a global and individual level. Finally, a secondary but not less important conclusion is the verification of the substantial similarity of the topologies of the virtual and physical components. This confirms the possibility, already stated elsewhere (Baggio et al., 2010a) of using the websites' network as a significant sample for the analysis of a tourism destination, which might greatly ease the data collection thus helping a growth in the application of network science in the study of tourism systems.

Although this study helps filling a gap in the existing literature and does offer some interesting implications for practitioners, it does have some limitations. In particular, the analysis of a narrow number of cases could be seen as a constraint on the outcomes presented here. However, the rigorous methodology employed coupled with the vast literature stating the crucial importance of ICTs in the tourism field, and other more general considerations on the validity of this type of case-study research (Flyvbjerg, 2006) allow us to confidently pose our conclusions as a general conjecture. More and more extensive studies will be able to confirm (or disproof) what attained in this paper.

Finally, a more extensive and deeper discussion on the importance of the DBE concepts in tourism is definitely needed. Space constraints forbid us to do it here, but future work on this issue is already planned.